\documentclass[conference,compsoc]{IEEEtran}
\ifCLASSOPTIONcompsoc
  \usepackage[nocompress]{cite}
\else
  \usepackage{cite}
\fi
\ifCLASSINFOpdf
\else
\fi
\usepackage{amsmath}
\usepackage{graphicx} 
\usepackage{algorithmic}

\usepackage{array}
\usepackage{multirow}

\begin{document}
%
\title{Multi-source imagery fusion using deep learning in a cloud computing platform}


\author{\IEEEauthorblockN{Carlos Theran\IEEEauthorrefmark{1},
Michael Alvarez\IEEEauthorrefmark{2},
Emmanuel Arzuaga\IEEEauthorrefmark{1}\IEEEauthorrefmark{2} and
Heidy Sierra\IEEEauthorrefmark{1}}
\IEEEauthorblockA{Laboratory for Applied Remote Sensing, Imaging and Photonics}
\IEEEauthorblockA{\IEEEauthorrefmark{1}Department of Computer Science \& Engineering}
\IEEEauthorblockA{\IEEEauthorrefmark{2}Department of Electrical \& Computer Engineering}
University of Puerto Rico Mayaguez
}

%


\maketitle

\begin{abstract}

Given the high availability of data collected by different remote sensing instruments, the data fusion of multi-spectral and hyperspectral images (HSI) is an important topic in remote sensing. In particular, super-resolution as a data fusion application using spatial and spectral domains is highly investigated because its fused images is used to improve the classification and tracking objects accuracy.
On the other hand, the huge amount of data obtained by remote sensing instruments represent a key concern in terms of data storage, management and pre-processing. This paper proposes a Big Data Cloud platform using Hadoop and Spark to store, manages, and process remote sensing data. 
Also, a study over the parameter \textit{chunk size} is presented to suggest the appropriate value for this parameter to download imagery data from Hadoop into a Spark application, based on the format of our data.
We also developed an alternative approach based on Long Short Term Memory trained with different patch sizes for super-resolution image. This approach fuse hyperspectral and multispectral images. As a result, we obtain images with high-spatial and high-spectral resolution. The experimental results show that for a chunk size of 64k, an average of 3.5s was required to download data from Hadoop into a Spark application. The proposed model for super-resolution provides a structural similarity index of 0.98 and 0.907 for the used dataset.             
\end{abstract}


%
\IEEEpeerreviewmaketitle

\section{Introduction}
Nowadays, many satellites that capture different information from objects have been developed to study Earth's surface features such as vegetation, rock formations, soil, water, snow, and human structures; generating huge amounts of data.
For instance, the Airborne Imaging Spectrometer (AVIRIS) has been used in a large number of experiments and field campaigns \cite{GREEN1998227}, the Hyperion instrument on board National Aeronautics and Space Administration (NASA)’s Earth Observing One (EO-1) spacecraft \cite{Ungar2003}, and Compact High Resolution Imaging Spectrometer (CHRIS) on ESA’s Proba-1 microsatellite \cite{Barnsley2004}, among others. 
These satellites capture images daily in various domains, for example, images with high spatial resolution, high spectral resolution, or temporal resolutions. 

In this manner, remote sensing is defined as big data problem, following the 5Vs definition of Big Data (volume, variety, velocity, veracity, and value) \cite{Chi2016}, carrying new challenges on data storage, data management, and data processing. In order to overcome the challenges related to data storage and management, researchers have proposed some parallel and distribute techniques using super-computers \cite{Zhenlong2018,Muelder2009,Judd1998,Benjamin2010,Cappello2005}. However, cloud computing technology has gained a lot more of attention due the advantage of commodity computer and storage devices. Its popularity has increased for the big pool of resources offered to users with low-cost, high-availability, scalability, storage, and computing power. For example, cloud computing platforms such as Google Cloud, Amazon Web Service, and Azure are provided by computer and software giant companies such as Google, Amazon, and Microsoft, respectively. The major advantages of cloud computing technology on big distributed data processing are: high reliability provided by fault tolerance mechanisms, scalability by virtualization technology, easy parallel programming, low cost on storage, and computing devices \cite{Wang2013}. In addition, the data processing problem has been tackled by using machine learning techniques to analyze, discriminate and classify information from the given data.

Thus, this paper proposes to use a cloud-based platform to manage remote sensing big data using Hadoop Distributed File System (HDFS) \cite{Shvachko2010}. Hadoop splits data sets (files) into chunk and distributes them across commodity computers called nodes, which are connected to each other and work together as a single system. Then, to analyze and process the data, a special node called client node can be used to access this data using different SparkSQL queries or MapReduce functions. Spark is a well known big data tool used in cloud computing processing. It is composed of a set of records or subjects of specific types, in which the data is partitioned and distributed across multiple nodes in the cluster. The main property of Spark is the Resilient Distributed Dataset (RDD) has the ability to store the data in the memory of each node, and makes it possible to process the data in parallel \cite{Zaharia2012,Zaharia2010}. In this work, a cloud computing platform for remote sensing storage, management, and  processing is developed using Hadoop and Spark.  

Also, a different approach based on Long Short Term memory is implemented to fuse Hyperspectral and Multispectral images. As a result, images with high-spacial and high-spectral resolution is obtained. The proposed approach is trained using different patch sizes preventing the spatial loss, and intrinsically provides data augmentation strategy. We use the structural similarity index, and signal to noise ratio to evaluate the quality of the fused images.

This paper is organized as follows:  Section 2 provides the methodology use for cloud computing and the proposed fusion method, Section 3 presents the description of the data, Section 4 presents experimental procedure, Section 5 presents the experimental results and, section 6 conclusion.



\section{Methodology}
This section provides the cloud computing platform's configuration details as a resource for management, storage, and processing data fusion task in remote sensing imaging.  Moreover, this section describes how HSI and MSI are formatted to be supported by HDFS. In this manner, we can overcome the management and storage challenges in the imagery area. In addition, a new approach for HSI and MSI fusion based on Long Short Term Memory (LSTM) is presented. This fusion approach is used as a test base to provide the performance of the proposed cloud configuration. 


\subsection{Hadoop Distributed File System}
HDFS is a particular distributed file system for efficient management and storage of massive structured and unstructured data. Also, It is designed to run on commodity hardware \cite{Shvachko2010}. The popularity of HDFS is highly fault-tolerant and reliable access to the data throughout applications. Our cloud configuration consists of HDFS master/slave architecture. The Master or Namenode manages the operations related to the metadata and filesystem namespace and regulates access to files by clients. Such operations are closing, opening, and renaming files and directories.
Meanwhile, the slaves or Datanodes are in charge of storing the chunks generated by the file splitting. In addition, It is responsible for reading and writes requests made by the client through the Namenode. Also, Datanodes perform chunks creation, deletion, and replication against a request from the NameNode. 

Our cloud configuration uses Apache YARN (Yet Another Resource Negotiator) that provides support for computing distributed paradigms, which was built focus on strong fault tolerance for massive, data-intensive computation \cite{Vavilapalli2013}. YARN works through APIs that request and works with cluster resources; consequently, it is the Hadoop's cluster resource management system. YARN application can allocate resources in the cluster by making all of its requests upfront or dynamically requesting the resources. In our case, an upfront approach is adopted because our application is based on Apache Spark \cite{Wu2016}, which is discussed later. Spark can be deployed in two different modes for running on YARN: YARN client mode, where the driver runs in the client, and YARN cluster mode, where the diver runs on the cluster in the YARN application master. Since we decide to have the opportunity to debug the output of our programs immediately, The YARN client mode was selected as a part of our cluster configuration. On the other hand, using YARN cluster mode, we are not able to have the interact component, such as \textit{spark-shell} for scala or java applications and \textit{pyspark} for python applications. Also, launching Spark application in client mode, the driver runs in the client process, and the application master is only used for requesting YARN resources. As a result, we are not overloading our cloud infrastructure if many clients request services on the cloud.

\subsection{Apache Spark}
\begin{figure}[h]
        \centering
         \includegraphics[width=0.4\textwidth,trim={0cm 0cm 0cm 0cm},clip]{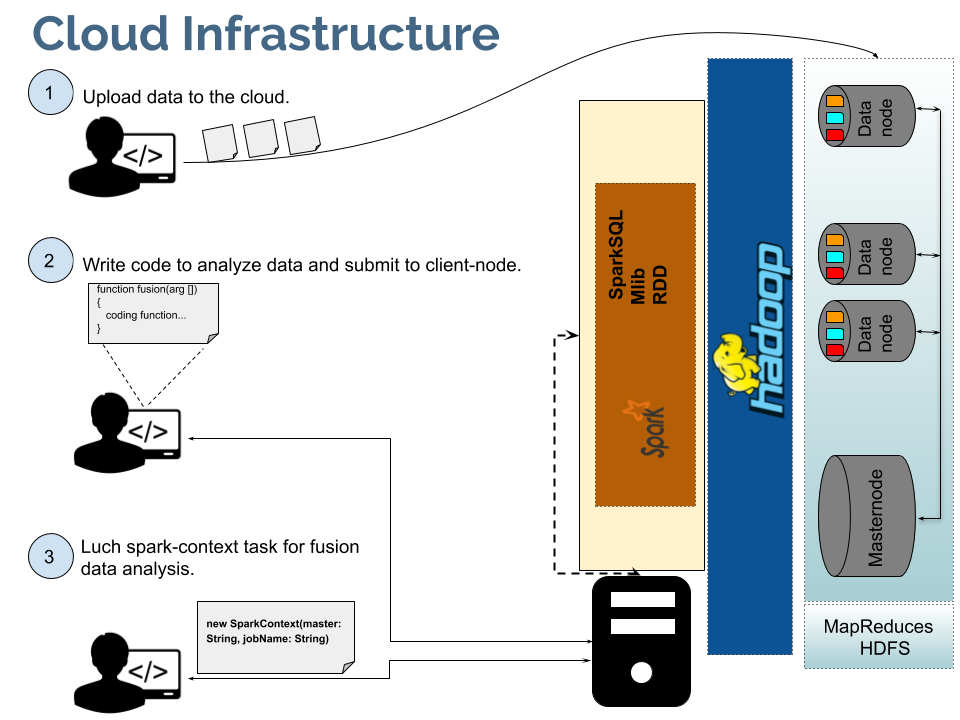} 
         \caption{Cloud infrastructure}
         \label{fig:cloud_infrastructure}
\end{figure}

Apache Spark is a cluster computing framework that introduces an abstraction called resilient distributed datasets (RDDs). An RDD is a read-only collection of objects partitioned across a set of machines that can be rebuilt if a partition is lost \cite{Zaharia2010,Zaharia2012}. Spark is well known for its ability to load large working dataset in memory. Consequently, it can improve the performance time of our application. The literature has demonstrated that Spark is the most appropriate API to be integrated within Hadoop \cite{Gu2013,Mavridis2017}. For example, Gu shows that Spark can achieve better performance in response time since the in-memory access over the cluster's distributed machines. As a result, Spark is faster than Hadoop in iterative operations, but a penalty in memory consumption is paid \cite{Gu2013}. Also, applications running as standalone on Spark have demonstrated to be faster than using only Hadoop \cite{Mavridis2017}.On the other hand, Spark can work with DataFrame as well. In this case, Spark uses a module named Spark SQL for structured data processing. DataFrame is a structured collection of data organized into named columns, constructed from structured data files, tables, external databases, or existing RDDs. It follow then, the proposed data fusion for super-resolution makes use of these two types of data structure. Mainly, This RDD and DataFrame are responsible for reading and write the HSI into our HDFS. Figure \ref{fig:cloud_infrastructure} presents a our cloud infrastructure.  

\subsection{Data Fusion for Super-resolution HSI}
Data fusion is becoming the preferred option to improve the data collected by multi-sources. As a result, we can achieve inferences that are not obtaining from a single source. Hence, satellite imagery has adopted data fusion techniques for the high availability of data in varied resolution domains (spatial, spectral, and temporal) of the same spacial scene, and an improvement over these resolution domains can be performed. Particularly, due to the high spectral content of HS image, it has gained relevant attention in remote sensing for image enhancement using data fusion techniques \cite{Ghamisi19}. However, typical hyperspectral sensors generate images with a high spectral resolution while sacrificing spatial resolution. Then, to overcome the lack of spatial resolution from the HSI, it is fuse with MSI due to its high spatial resolution. This technique has gained relevant attention to generate images with high-spatial and high-spectral resolution (HSaHSe) \cite{Theran2019,Palsson2017}.

\begin{figure}[h]
        \centering
         \includegraphics[width=0.5\textwidth,trim={0cm 0cm 0cm 0cm},clip]{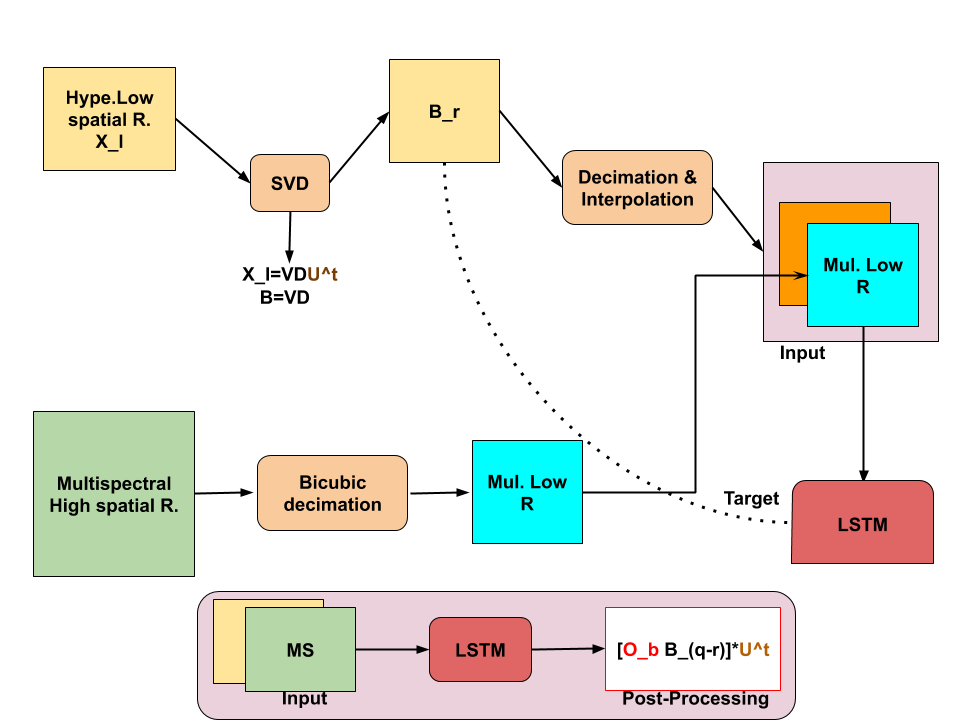}
         \vspace{-0.2cm}
         \caption{Scheme fusion for super-resolution.}
         \label{fig:schema fusion}
\end{figure}

In this paper we provide a alternative approach base on the well known  LSTM \cite{Hochreiter1997} to fuse HSI and MSI. The LSTM network is the inclusion of a self-loop that avoid the exploiting gradient and vanishes gradient problem presented in other networks \cite{Goodfellow2016}. This network use a set of parameters to control the shared information through different hidden layers, these parameters are defined as follows for a time $t\in\mathbf{N}$.  
\[ \begin{pmatrix} 
   f_i \\
   i_i \\
   \hat{c}_i \\
   c_i \\
   o_i \\
   h^{t} \\
\end{pmatrix} =\small{
  \begin{cases}
    \sigma\Bigg( b_i^f + \sum_j U_{i,j}^f x_j^{(t)} + \sum_j W_{i,j}^f h_j^{t-1}\Bigg)\\
    \sigma\Bigg( b_i + \sum_j U_{i,j} x_j^{(t)} + \sum_j W_{i,j} h_j^{t-1}\Bigg)\\
    \tanh\Bigg( b^c_i + \sum_j U_{i,j}^c x_j^{(t)} + \sum_j W_{i,j}^c h_j^{t-1}\Bigg)\\
    f^{(t)}_i \times c_i^{(t-1)}+ i^{(t)}_i \times \hat{c}_i^{(t)} \\
    \sigma\Bigg( b^o_i + \sum_j U_{i,j}^o x_j^{(t)} + \sum_j W_{i,j}^o h_j^{t-1}\Bigg)\\
    \tanh(c_i^t)o_i^t
  \end{cases}}
\]

where $x^{(t)}$ is the actual input vector, $\textbf{h}^{(t)}$ is the output of the LSTM at the current hidden layer, and $\textbf{b}^{(f,c,o)}$, $\textbf{U}^{(f,c,o)}$, $\textbf{W}^{(t,c,o)}$ are biases, input weights, and recurrent weights, respectively. The proposed approach uses the LSTM to propagate the learned spatial information through given sequences of different patch sizes, which are taken from the same spatial scene. The proposed approach is divide in three different steps; Initially a data prepossessing is required to separate the load content and the spectral content from the HSI, which has a low-spatial resolution. Also, a quinctic decimation over the MSI is perform. The two results of these processes are concatenated to generate the result of the first step. Secondly, an LSTM model is trained using different blocks size to generate an image with high spatial resolution. Finally, the spectral content extracted in the first step is multiply with the output of the LSTM.  Figure \ref{fig:schema fusion} shows the workflow of the super-resolution process fusing HSI and MSI. 

\section{Data Description}
\begin{figure}[h]
\centering
\includegraphics[width=.4\textwidth]{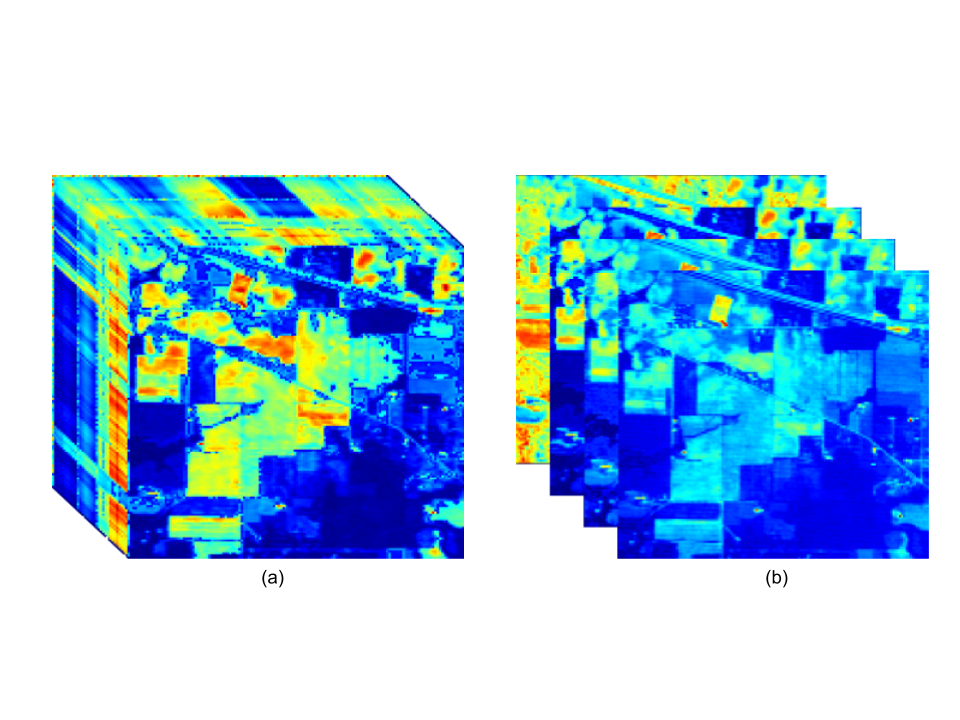}
\vspace{-1.2cm}
\caption{Figure (a) is the HSI image of Indian and figure (b) the MS image.}
\label{fig:Indian}
\end{figure}

The experimental data used in this work consist in two data set, Indian Pines hyperspectral image and Enrique reef hyperspectral image. Indian Pines had the following characteristic. This image was gathered by the AVIRIS sensor and consists of $145\times 145$ pixels and 224 spectral bands in the wavelength range 400 to 2500 nm. The number of bands were reduced to 200 by removing high water absorption bands. This scene has 16 classes: Alfalfa, Corn-notil, Corn-mintil, Corn, Grass-pasture, Grass-trees, Grass-pasture-mowed, Hay-windrowed, Oats, Soybean-notill,  Soybean-mintill, Soybean-clean, Wheat, Woods, Buildings-Grass-Trees-Drives, and Stone-Steel-Towers.
See figure \ref{fig:Indian}.
\vspace{-1 cm}

\begin{figure}[h]
\centering
\includegraphics[width=.4\textwidth]{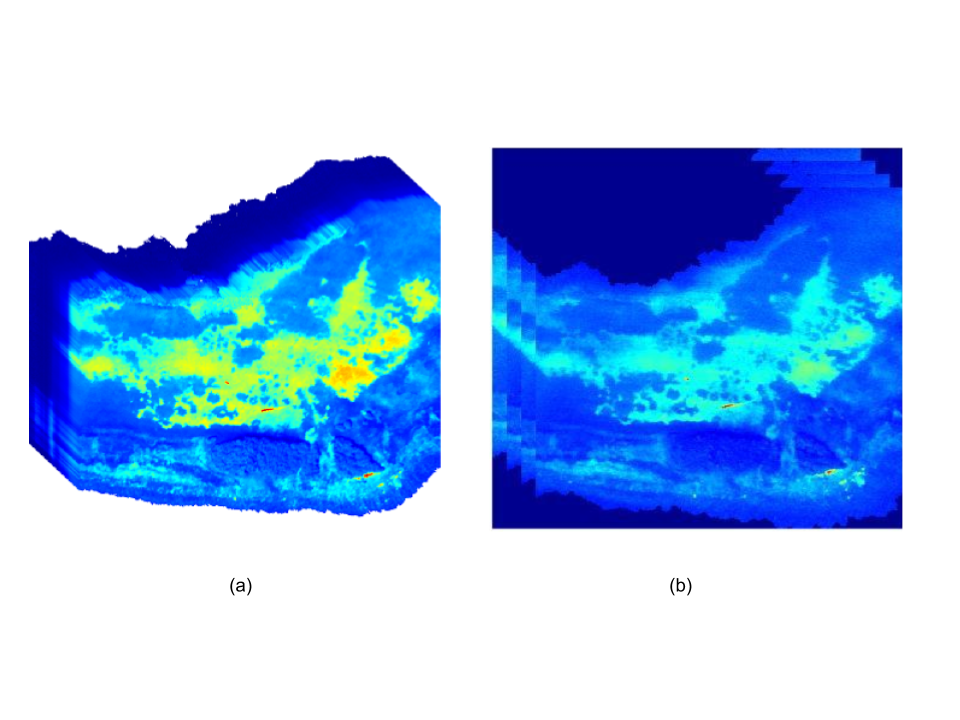}
\vspace{-1.2cm}
\caption{Figure (a) is the HSI image of Enrique Reef and figure (b) the MS image.}
\label{fig:Enrique}
\end{figure}

 The Enrique Reef hyperspectral images consists 102 bands taken from the from the AISA Eagle sensor, this image was acquired in 2007.  The spatial resolution of this data is 1m.There are 6 classes: Mangrove, Deep water, Coral, Sand, Sea grass, and Flat reef. See figure \ref{fig:Enrique}
 
These images are used to generate simulated data using the procedure presented in \cite{Palsson2017,Theran2019}. As a results, a low spatial resolution hyperspectral image is simulated by applying image decimation by a factor of 4 to the original dataset. Likewise, a high spatial resolution multispectral image is simulated from the original data set by averaging bands from different spectral ranges: 1) Blue: 445-516nm, 2) Green: 506-595nm, 3) Red: 632-698nm and 4) NIR = 757-853nm.

\subsection{Data Preprocessing}
\begin{figure}[h]
\centering
\includegraphics[width=.5\textwidth]{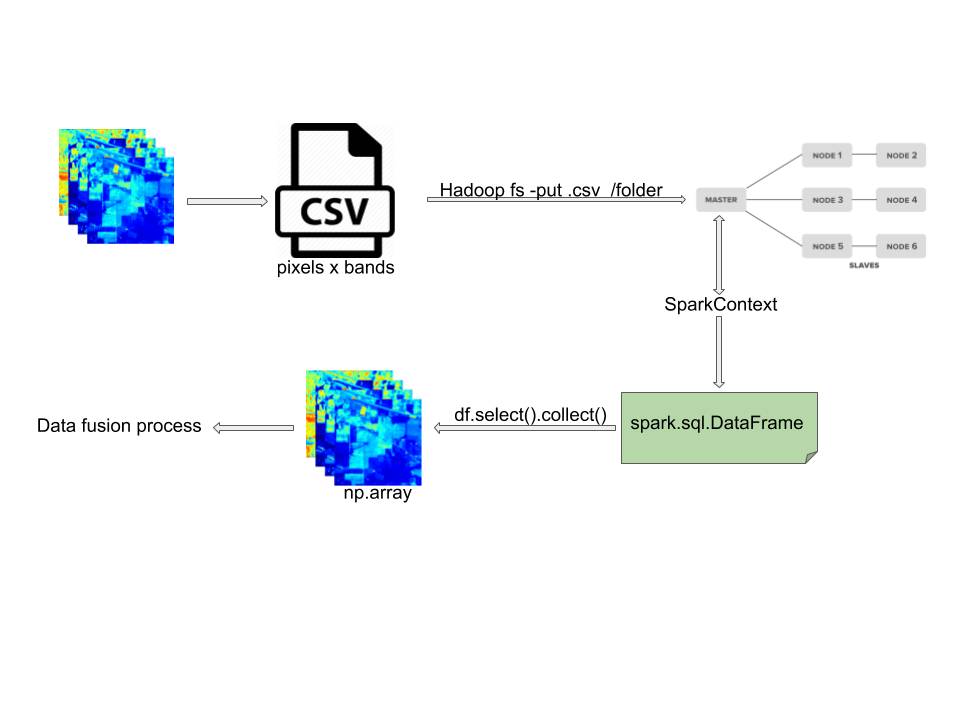}
\vspace{-2.5 cm}
\caption{Schema for data transformation.}
\label{fig:transformation}
\end{figure}

Different transformations over the data presented in figure \ref{fig:transformation} were required to be able to execute the schema presented in figure \ref{fig:schema fusion} on a cloud environment. Spark is an engine that uses two different data types, \textit{RDD} and \textit{spark.sql.DataFrame}, As a consequence, we need to transform the data to be load as one of those two data types. 

Figure 5 describes the data transformation process to load the images into an spark application using spark.sql.DataFrame data type.

\section{Experimental Procedure}
The cloud computing environment for the presented experiments consists of one cluster configuration base on Hadoop and Spark. To provide the following results, we built a prototype that is scalable using more hardware resources. In our case, six virtual machines (VM) were created; five of these VMs are using 10 GB of storage and 6 G DDR4, and one has 10 GB and 16 G DDR4. The operating system is Ubuntu 16.04 LTS and each machine have 4 VCPU. 

In the following experiment, a variation of the chunk size occurs to analyze the time required to download the remote sensing data from Hadoop into an application using Spark. The tuning of this parameter allows us to identify the correct chunk size to improve the application's execution time. In \cite{Kang2012} present a description of the chunk size for remote sensing images, but do not provide experimental results with real data set. The chunk size considers for these experiments are 4KB, 16K, 32K, 64K, 128k, 1M, 10M, and 100M. We are taking values below to the default in Hadoop 128M \cite{Shvachko2010}, to avoid the needs of more memory for sort map task output, which can crash the Java Virtual Machines or add significant garbage collection overhead.

On the other hand, the HSaHSe generated by our method presented in figure \ref{fig:schema fusion} will be test using SSIM, and PSNR metrics \cite{wang2004,Fallah2010}, these metrics have been used in recent publication in HS and MS fusion \cite{Palsson2017,Theran2019}:

\begin{equation}
\label{eq:ssim}
 SSIM(x,y) = \frac{(2\mu_x\mu_y+c_1)(2\sigma_{xy}+c_2)}
                  {(\mu_x^2+\mu_y^2+c_1)(\sigma_x^2+\sigma_y^2+c_2)}
\end{equation}

\noindent in equation \eqref{eq:ssim}, $\mu_x$ and $\mu_y$ are the average of $x$ and $y$, respectively. $\sigma_x^2$ and $\sigma_y^2$ are the variance of $x$ and $y$ respectively, and $c_1=(k_1,L)^2$, $c_2=(k_2,L)^2$, where $k_1=0.01, k_2=0.03$, and $L$ is the dynamic range of the pixel-values. The PSNR is modeled by
\begin{equation}
    PSNR = 10\log\Big(\frac{R^2}{RMSE}\Big)
\end{equation}
\noindent 
where $RMSE$ is the well known Root Mean Square Error formula, and $R$ is the maximum fluctuation in the input image. A high SSIM means, the images generated by the fusion of the HSI and MSI have high similarity to the reference images in structural information. Along the same line, PSNR compares the level of the desired signal to the level of background noise. Thus, a higher PSNR means that there is more useful content in the obtained data.

The algorithm based on the schema in figure \ref{fig:schema fusion} was implemented in python. For the training phase the packets required are Tensorflow 2.2, and Keras 2.0.     

\section{Experimental Results}
This section presents the performance of our cloud infrastructure in terms of the time needed to download our files from Hadoop into our Spark application. Also, the results of the proposed method for super-resolution fusing HSI and MSI is presented.
\subsection{Cloud Computing Metrics}
\begin{figure}[h]
\centering
\includegraphics[width=.5\textwidth]{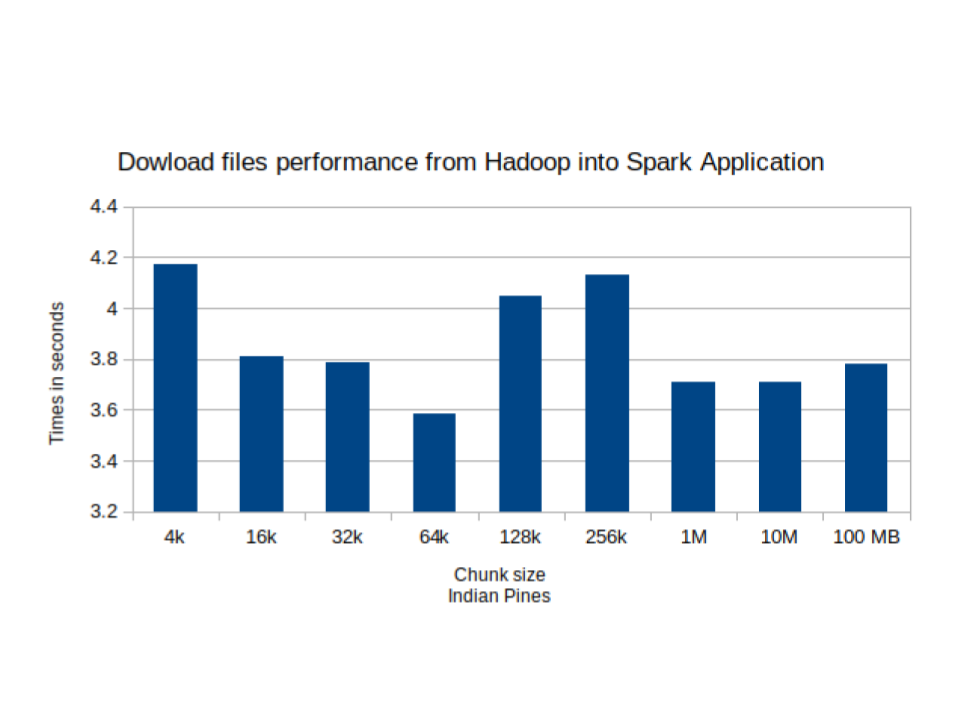}
\vspace{-1.8 cm}
\caption{Study of chunk size parameter for HSI and MSI Indian pines data.}
\label{fig:chuck_size_metric_indian}
\end{figure}

\begin{figure}[h]
\centering
\includegraphics[width=.5\textwidth]{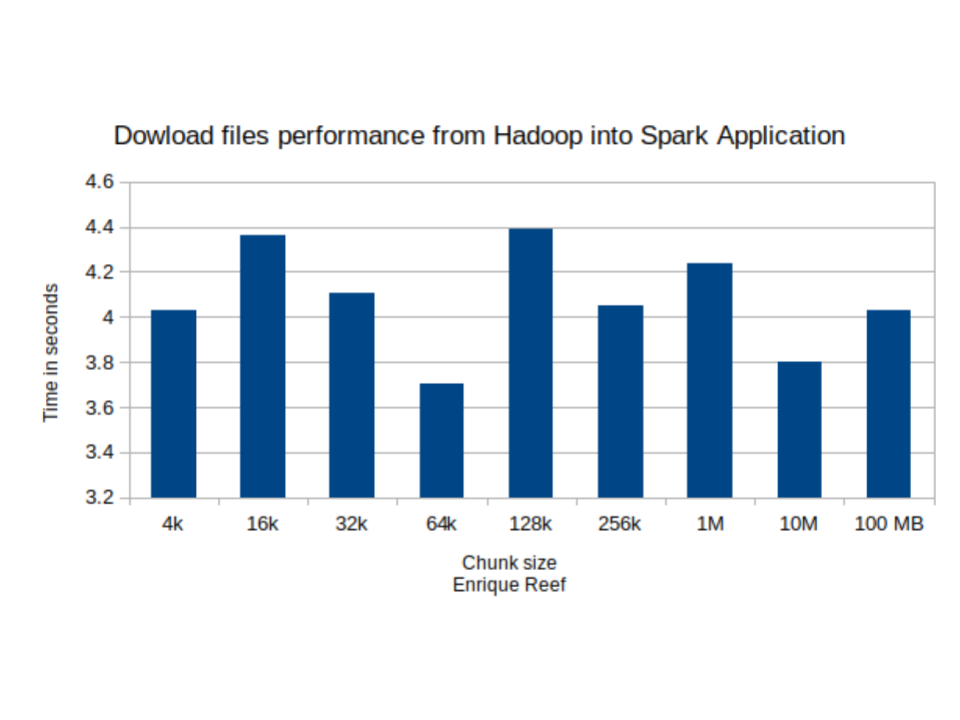}
\vspace{-1.8 cm}
\caption{Study of chunk size parameter for HSI and MSI Enrique Reef data.}
\label{fig:chuck_size_metric_enrique}
\end{figure}

To take the best advantage of the cloud platform, we studied the download performance from Hadoop to a Spark application in terms of the parameter chunk size, in particular, for the proposed data fusion application and using the data format presented in figure \ref{fig:transformation}. The correct value for this parameter can provide an improvement in the execution time of the spark application. In this experiment the set of chunk size described in section 4 was selected. Figure \ref{fig:chuck_size_metric_indian} and \ref{fig:chuck_size_metric_enrique} show the time in seconds spend to download the data from Hadoop into an Spark Application. For both data set different performances was obtain. Even though, 64k can be consider a good parameter for both data set. Figure  \ref{fig:chuck_size_metric_indian} presents the results using Indian Pines image, and figure \ref{fig:chuck_size_metric_enrique} shows the results for Enrique Reef. These results were computed using the average of 10 execution of the super-resolution Spark application.      

Figure \ref{fig:chuck_size_metric_indian} shows that for 64k the running time is lower with 3.5s. It means, the acquisition of the data form Hadoop into Spark application is faster for these values. Now, the standard deviation for the 90 execution was 0.58.

Also, Figure \ref{fig:chuck_size_metric_enrique} shows that for 64k the running time is lower with 3.7s. As a result, for 64k a fast acquisition of data from Hadoop is performance. Now, the standard deviation for the 90 execution was 0.50.

\subsection{Super-resolution HSI Results}
For experimental propose the following sets of patches size to train our model to generate HSaHSe image were selected; $\Lambda_1=\{16\times 16, 14\times 14,12\times 12,10\times 10\}$, $\Lambda_2=\{12\times 12, 10\times10,8\times8,6\times 6\}$, $\Lambda_3=\{8\times 8, 6\times6,4\times4,2\times 2\}$. For each set, the following number of patches were used to train the model in order to obtain an $80\%$ of data training. For $\Lambda_1$, 179 and 1772 pathch were used in Indian Pines and Enrique Reef, respectably.  For $\Lambda_2$ were used $272$ and $3021$ in Indian Pines and Enrique Reef, respectably. And $562$, $6750$ for Indian Pines and Enrique Reef in $\Lambda_3$. Table \ref{tb:summaryIndian} and \ref{tb:summaryEnrique} provide the results of the proposed model for HSI and MSI fusion.  

\begin{center}
\begin{table}[]
\centering
\begin{tabular}{|c|cl|cl|}
\hline
\multirow{2}{*}{Patch size}                & \multicolumn{2}{c|}{SSIM}                            & \multicolumn{2}{c|}{PSNR}                            \\ \cline{2-5} 
                                  & \multicolumn{1}{c|}{$\bar{x}$}   & \multicolumn{1}{c|}{$\sigma$ }   & \multicolumn{1}{c|}{$\bar{x}$}   & \multicolumn{1}{c|}{$\sigma$ }   \\ \hline
$\Lambda_1$              & 0.907        & 3.6E-03    & 29.99        & 8.8E-01    \\
$\Lambda_2$                & 0.906        & 1.1E-03    & 29.98        & 2.7E-02    \\
$\Lambda_3$                 & 0.906        & 1.5E-03    & 29.97        & 7.1E-02 \\  \hline
\end{tabular}
\vspace{0.2 cm}
\caption{\small{Summary of the proposed technique using Indian Pines image. An 80\% training samples were used. The presented values were the best performance of the proposed model combining using different patch size.} }
\label{tb:summaryIndian}
\end{table}
\end{center}

\begin{figure}[h]
\centering
\includegraphics[width=.5\textwidth]{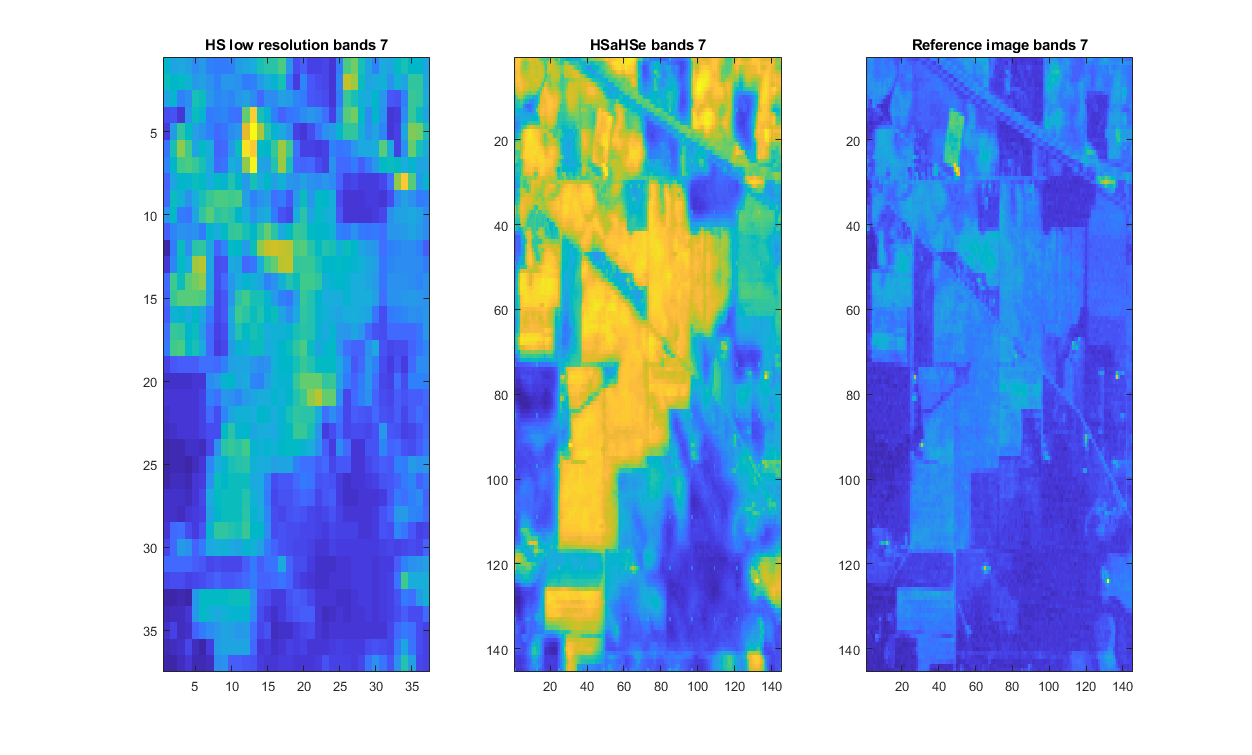}
\vspace{-1.0 cm}
\caption{From left to right; the HSI low spatial resolution, the HSaHSe and the reference image.The band 7 was selected in this figure.}
\label{fig:band_7}
\end{figure}

\begin{figure}[h]
\centering
\includegraphics[width=.5\textwidth]{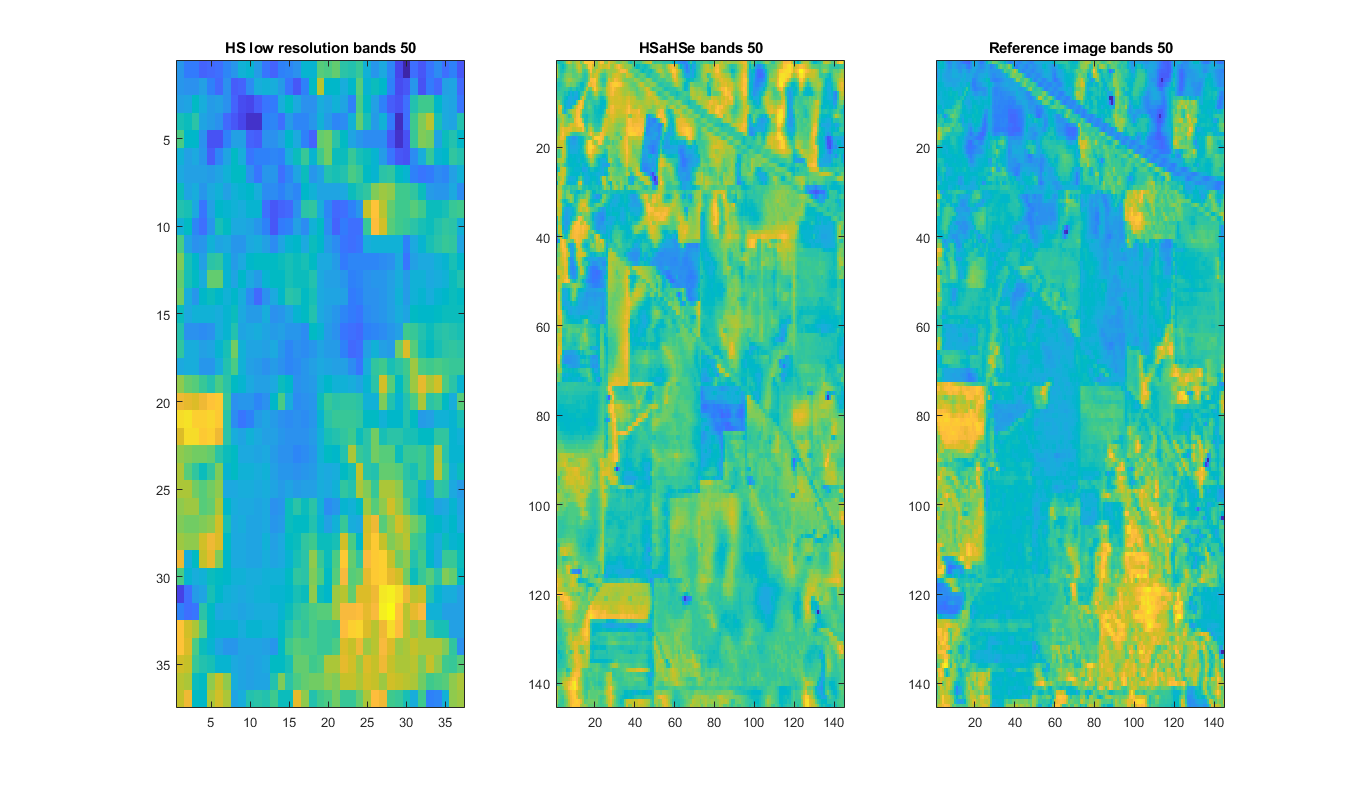}
\vspace{-1.0 cm}
\caption{From left to right; the HSI low spatial resolution, the HSaHSe and the reference image. The band 50 was selected in this figure.}
\label{fig:band_50}
\end{figure}

\begin{figure}[h]
\centering
\includegraphics[width=.5\textwidth]{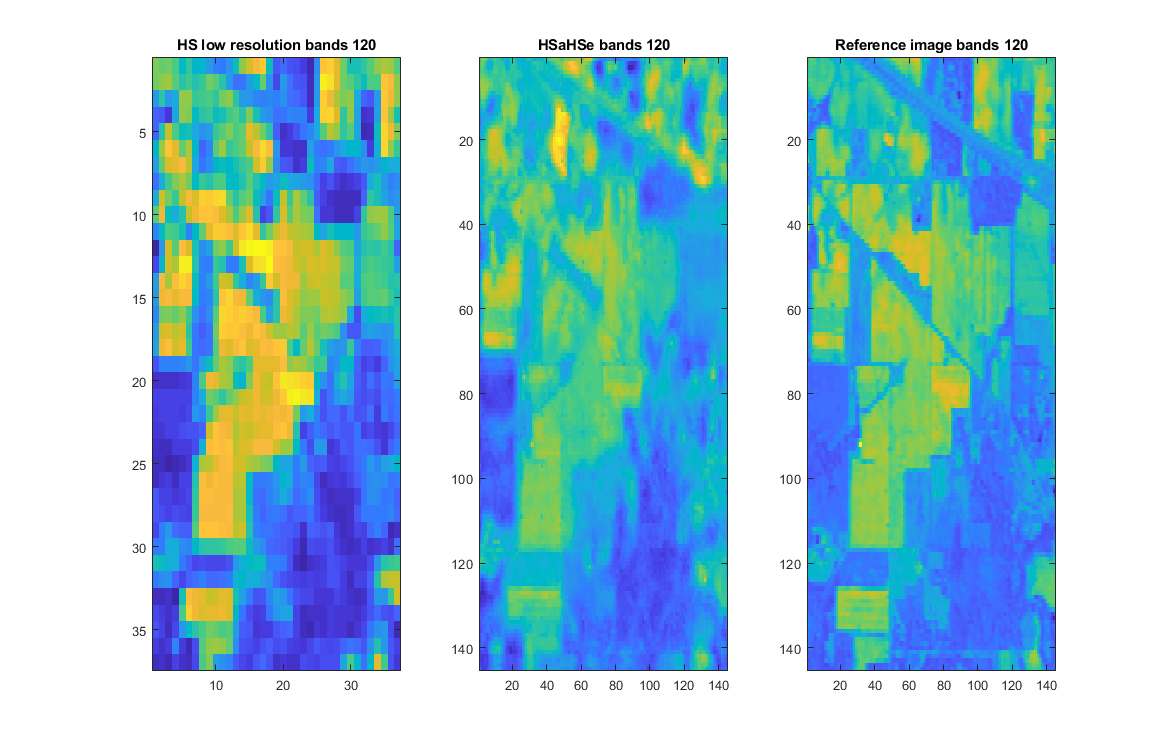}
\vspace{-1.0 cm}
\caption{From left to right; the HSI low spatial resolution, the HSaHSe and the reference image. The band 120 was selected in this figure.}
\label{fig:band_150}
\end{figure}

The table \ref{tb:summaryIndian} presents the SSIM obtained with the best case for the set of patches size $\Lambda_1$. Using this set, the SSIM was $0.907$, and the PSNR was $29.99$. It means that the proposed method provides an excellent reconstruction for HSI with a high spatial resolution with a low nose into the images. In addition, figures \ref{fig:band_7},\ref{fig:band_50}, and \ref{fig:band_150} show from left to right the HSI in low spatial resolution, the HSI generated by the proposed model, and reference image. 

From images \ref{fig:band_7},\ref{fig:band_50}, and \ref{fig:band_150}, we can observe that the proposed method is providing an excellent performance for HSI enhancement in the spatial content. 

\begin{center}
\begin{table}[]
\centering
\begin{tabular}{|c|cl|cl|}
\hline
\multirow{2}{*}{Patch size}                & \multicolumn{2}{c|}{SSIM}                            & \multicolumn{2}{c|}{PSNR}                            \\ \cline{2-5} 
                                  & \multicolumn{1}{c|}{$\bar{x}$}   & \multicolumn{1}{c|}{$\sigma$ }   & \multicolumn{1}{c|}{$\bar{x}$}   & \multicolumn{1}{c|}{$\sigma$ }   \\ \hline
$\Lambda_1$ & 0.987 & 1.6E-03    & 39.88        & 0.03 \\ \hline
$\Lambda_2$ & 0.987 & 2.4E-03    & 39.91        & 0.02 \\ \hline
$\Lambda_3$ & 0.987 & 9.5E-03    & 39.92        & 0.06 \\ \hline
\end{tabular}
\vspace{0.2 cm}
\caption{\small{Summary of the proposed technique using Enrique Reef image. An 80\% training samples were used. The presented values were the best performance of the proposed model combining using different patch size.} }
\label{tb:summaryEnrique}
\end{table}
\end{center}

The table \ref{tb:summaryIndian} presents the SSIM obtained with the best case for the set of patches size $\Lambda_3$. Using this set, the SSIM was $0.987$, and the PSNR was $39.92$. It means that the proposed method provides an excellent reconstruction for HSI with a high spatial resolution with a low nose into the images. Figures \ref{fig:band_7}, \ref{fig:band_50}, and \ref{fig:band_150} show from left to right the HSI in low spatial resolution, the HSI generated by the proposed model, and reference image. 

\begin{figure}[h]
\centering
\includegraphics[width=.5\textwidth]{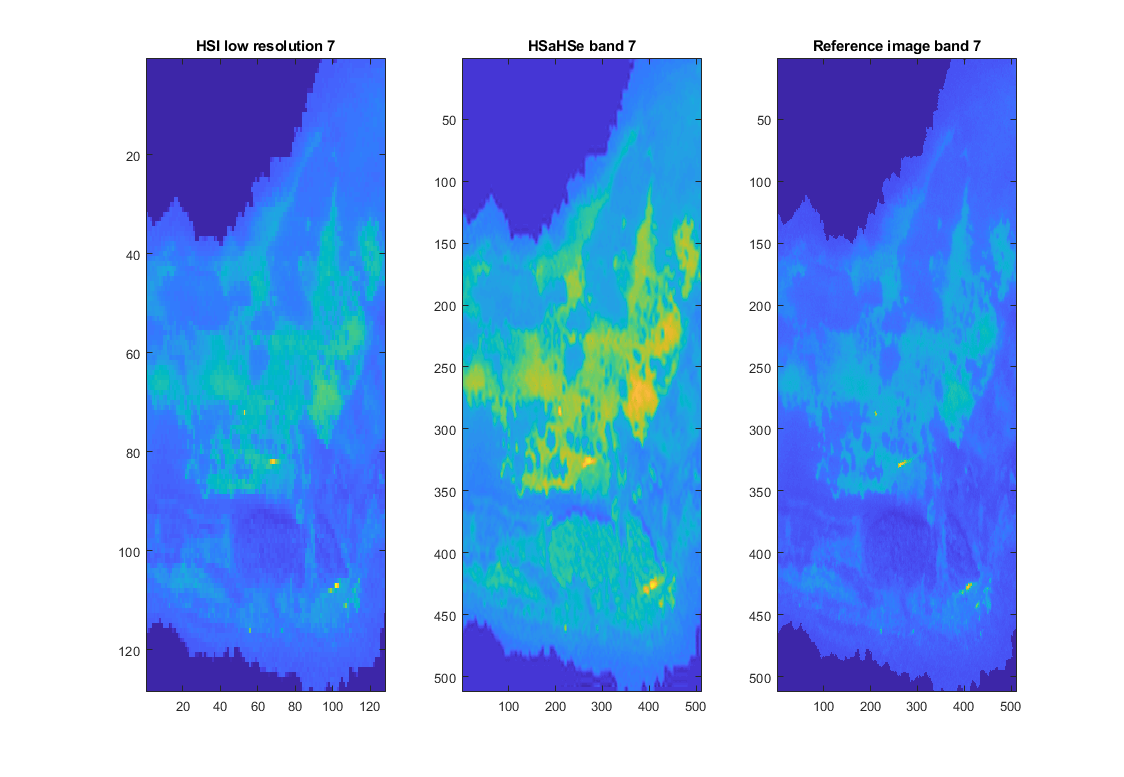}
\vspace{-1.0 cm}
\caption{From left to right; the HSI low spatial resolution, the HSaHSe and the reference image.The band 7 was selected in this sample.}
\label{fig:band_7_enrique}
\end{figure}

\begin{figure}[h]
\centering
\includegraphics[width=.5\textwidth]{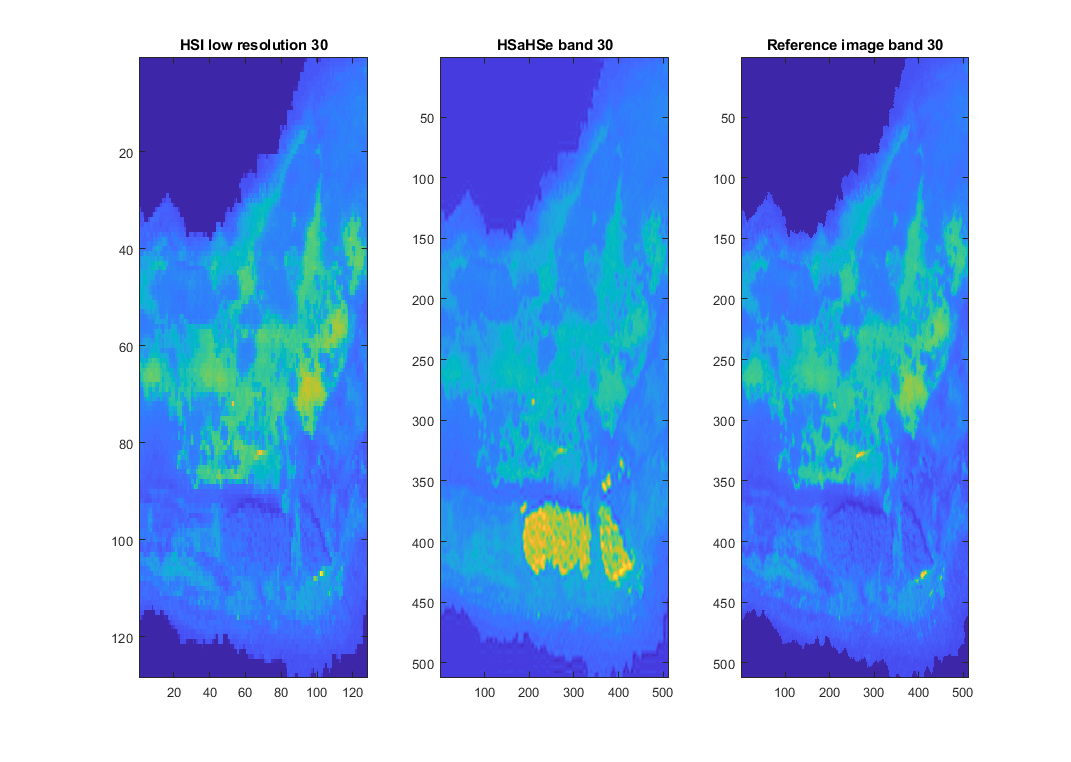}
\vspace{-1.0 cm}
\caption{From left to right; the HSI low spatial resolution, the HSaHSe and the reference image. The band 30 was selected in this sample.}
\label{fig:band_30_enrique}
\end{figure}

\begin{figure}[h]
\centering
\includegraphics[width=.5\textwidth]{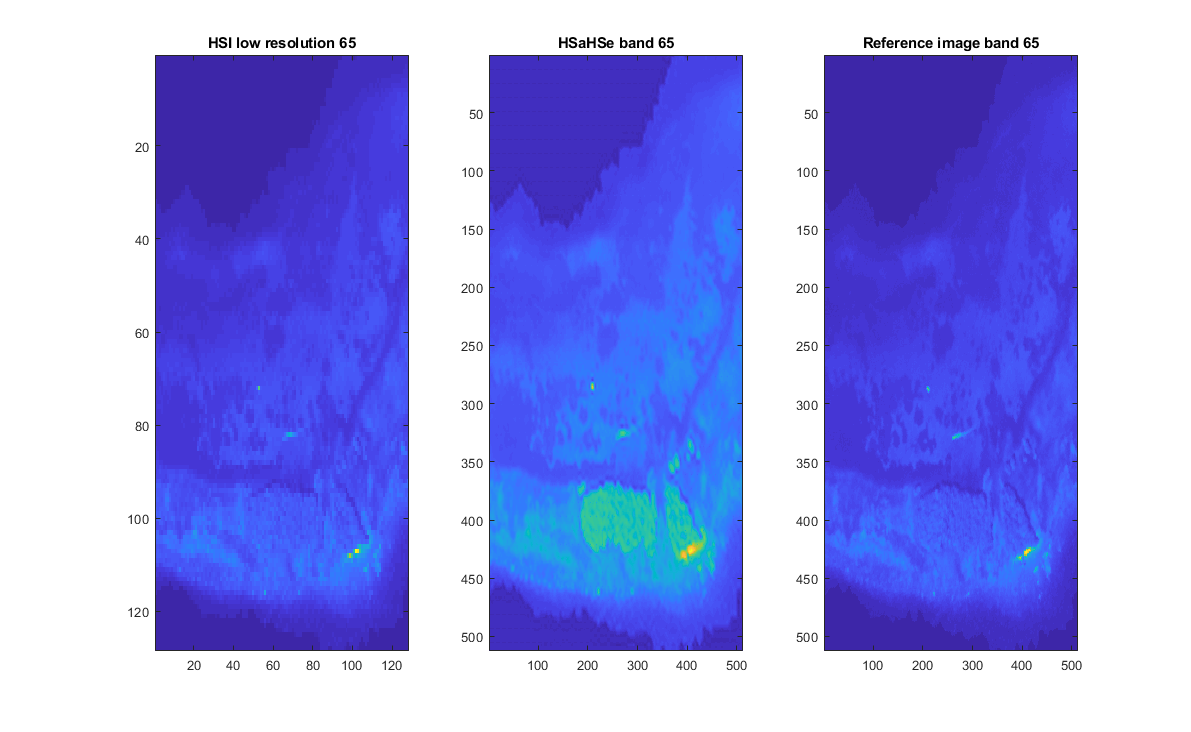}
\vspace{-1.0 cm}
\caption{From left to right; the HSI low spatial resolution, the HSaHSe and the reference image. The band 65 was selected in this figure.}
\label{fig:band_65_enrique}
\end{figure}

The images \ref{fig:band_7_enrique},\ref{fig:band_30_enrique}, and \ref{fig:band_65_enrique} show the performance of the proposed method for Enrique Reef data set. It provides an excellent performance for HSI enhancement in the spatial content. 

\section{Conclusion}
In order to overcome the needs in the area of satellite imagery, a cloud computing environment was configured using Hadoop and Spark. As a result, we can store, manage, and process remote sensing data into a cloud platform with a high fault-tolerance mechanism. The parameter chunk size was studied to determine the chunk's best size to get the best benefits of Hadoop and Spark. 
In particular, for our data, the best chunk size is 64K. In addition, a new process HSI and MSI data transformation were presented to perform data fusion techniques into the Spark environment. Also, a new approach for HSI and MS fusion was presented. As a result, an HSaHSe image was generated with a 0.907 of SSIM and 29.97 of PSNR for Indian Pines data set, and 0.987 of SSIM and 39.88 of PSNR for Enrique Reef data set.

\ifCLASSOPTIONcompsoc
  \section*{Acknowledgments}
\else
  \section*{Acknowledgment}
\fi

The authors would like to thank M.Sc. Alejandro E. Gonzalez for the deployment of Virtual Machines where the cloud was built. This work is partially supported by NSF Grant No. OAC-1750970 and NSF Award No. OIA-1849243.



\bibliographystyle{IEEEtran}
\bibliography{references}
%



\end{document}